# Average energy approximation of the ideal Bose-Einstein gas and condensate

Don S Lemons

Bethel College

North Newton, KS 67206

dlemons@bethelks.edu

ABSTRACT

If the $N$ bosons that compose an ideal Bose-Einstein gas with energy $E$ and volume $V$ are each assumed to have the average energy of the system $E/N$, the entropy is easily expressed in terms of the number of bosons $N$ and the number of single-particle microstates $n$ they can occupy. Because the entropy derived is a function of only $N$ and $n$, and the latter is a function of the extensive variables, $E$, $V$, and $N$, this entropy describes all that can be known of the thermodynamics of this system. In particular, the entropy very simply recovers the Sakur-Tetrode entropy in the classical limit and at sufficiently low temperature describes an unstable system. A thermodynamic stability analysis recovers the Bose-Einstein condensate and a two-phase region. Apart from numerical factors of order one, results are identical with those derived via standard, probabilistic methods.

1. Introduction

The Bose-Einstein gas and its condensate [1] are standard topics in statistical mechanics and thermal physics texts [2, 3, 4, 5, 6, 7, 8, 9, 10, 11, 12, 13]. These texts model the Bose-Einstein gas by separating the gas particles into groups with a common temperature and various energies described by a density of states function. This approach employs the results of wave mechanics, a density of states function, probabilities of single-particle microstates, Lagrange multipliers, partition functions, and Riemann zeta functions. Numerical methods have been invented that explore the resulting equations of state [14]. The average energy approximation, according to which all the particles of an ideal Bose-Einstein gas have the same energy, dispenses with all this machinery.

Not only is the average energy model of the ideal Bose-Einstein gas and condensate simpler than the standard model it is alternative to it and, of course, it is always good to have different derivations of the same phenomena. But more importantly, the very minimalism of the average energy approximation isolates the essential ideas behind the ideal Bose-Einstein gas and condensate: the quantization of phase space into Planck's constant sized units, the indistinguishability of identical particles, and a careful choice for the number of single-particle microstates, and prevents these ideas from being lost in a forest of details.

The average energy approximation is part of a tradition that dates back, at least, to 1857 when Rudolf Clausius derived the ideal gas equations of state by assuming that all its particles have the same energy [15]. The basic tactic is that of mean field theory [16] in which a variable quantity is replaced with its average value before, rather than after, a critical calculation.

The average energy approximation as applied to the ideal Bose-Einstein gas is made possible by the insight that the number of single-particle microstates available to a single particle of ideal gas is independent of whether these single-particle microstates are filled with distinguishable or indistinguishable particles, and if indistinguishable whether bosons or fermions. Therefore, the number of single-particle microstates can be determined by partitioning the phase space of an N-particle ideal gas of distinguishable particles into units of action whose size is determined by Planck's constant. The best place to further explore the exact statement of the average energy assumption, its physical basis, and its *a priori* justification is in Section 3 of this paper.

The cost of not partitioning the ideal gas particles into many groups each with a common temperature and its own energy, that is, the cost of adopting the average energy approximation, is the loss of some accuracy. Thus, while the average energy approximation reproduces the expected phenomena and the standard functional dependences, some of the derived constants are different. For instance, the average energy approximation produces a critical temperature $0.130(N/V)^{2/3}(h^2/mk)$ at which the Bose-Einstein gas begins to condense that is 50% greater than the critical temperature $0.0839(N/V)^{2/3}(h^2/mk)$ derived with standard methods. Other comparisons with the results of the standard approach are noted in Section 10.

Nonetheless, a case can be made for the inclusion of the average energy model of the ideal Bose-Einstein gas and condensate in the undergraduate physics curriculum either relatively early as a way to introduce the physics of the ideal quantum gases or relatively late as a supplement to the standard approach. It may, in fact, be that the average energy approximation allows one to introduce the ideal quantum gases in an elementary course that builds closely on classical thermodynamics and the concept of statistical multiplicity whereas no such option existed before. More importantly, the aim of physics is not only to produce optimally accurate results that can be empirically tested but also to explore the relation of ideas, for instance, in the present case to distinguish between what ideas are required to bring the Bose-Einstein condensate into existence and what merely improve the accuracy of its prediction. Of course, it is important to clearly explain the nature of the approximations made and to remember that the standard approach also requires assumptions and issues in a model.

This paper consists of a sequence of short sections that lead the reader through the logic of the average energy approximation as applied to the ideal Bose-Einstein gas and condensate. Its important steps are: a review of the classical microstates available to an ideal gas of $N$ distinguishable particles (Section 2), the physical basis of the average energy approximation (3), the multiplicity of an ideal gas (4), the entropy of the ideal Bose-Einstein gas (5), its equations of state (6), its low temperature behavior (7), its thermodynamic instability (8), its two-phase region (9), and a summary and conclusion (10).

2. The Multiplicity of an Ideal Gas of Distinguishable Particles

Our immediate goal is to discover the multiplicity $\Omega$, sometimes called the degeneracy, of an isolated, ideal Bose-Einstein gas in terms of its extensive variables: internal energy $E$, volume $V$, and particle number $N$. Because in achieving this goal we borrow a key expression from classical statistical mechanics, we begin by reviewing the analysis of an ideal gas composed of distinguishable particles.

The most direct way of determining the multiplicity $\Omega(E,V,N)$ of an ideal gas of distinguishable particles is to imagine that the N-particle system occupies, with equal probability, positions in a 3N-dimensional volume $V^N$ and on a constant energy shell of radius $\sqrt{2mE}$ in a 3N-dimensional momentum space. In order to count the number of different positions in this phase space we divide the 3N-dimensional volume $V^N$ into uniform cells of size $\delta s^{3N}$ where $\delta s$ is a small but otherwise arbitrary length. We also discretize the system's momentum space by expanding the constant energy surface into a concentric spherical annulus of small but otherwise arbitrary thickness $\delta p$ and then dividing the volume of this annulus by the volume $\delta p^{3N}$ of a uniform cell in 3N-dimensional momentum space. Because the details of this calculation can be found in many texts [17, 18, 19, 20, 21, 22, 23, 24, 25, 26], we do not reproduce these details here.

One finds that the number of different phase space cells of size $(\delta s \delta p)^{3N}$ that can be occupied by an isolated, ideal gas composed of $N$ distinguishable particles with energy $E$ and volume $V$ is

$$\Omega(E,V,N) = \left[ V \left(\frac{E}{N}\right)^{3/2} \left(\frac{4\pi em}{3H^2}\right)^{3/2} \right]^N \quad (1)$$

where we have assumed that $N \gg 1$ and established the notation $H = \delta s \delta p$. Note that the symbol $e$ here, and throughout this paper, stands for the base of the natural logarithm. The dimensions of $H$ $[= \delta s \delta p]$ are those of action and the magnitude of $H$ is limited only by the requirement that $\Omega \gg 1$. The quantity $H$ is, in classical physics, an artifact that has no empirical consequences.

Each of the $\Omega(E,V,N)$ phase space cells determined in this way defines a classical microstate. Therefore, the multiplicity $\Omega(E,V,N)$ is simultaneously the number of phase space cells and the number of equally probable classical microstates that can be occupied by an N-particle, isolated, ideal gas of distinguishable particles when $N \gg 1$.

According to (1) the multiplicity $\Omega(E,V,N)$ is a factor

$$\Omega(E/N,V,1) = V \left(\frac{E}{N}\right)^{3/2} \left(\frac{4\pi em}{3H^2}\right)^{3/2} \quad (2)$$

raised to the power of $N$, that is,

$$\Omega(E,V,N) = \left[\Omega(E/N,V,1)\right]^N . \quad (3)$$

And if $\Omega(E,V,N)$ is the number of microstates available to a gas of $N$ distinguishable particles occupying volume $V$ and having energy $E$ when $N \gg 1$, then $\Omega(E/N,V,1)$ must be the number of microstates available to a single particle of gas occupying volume $V$ and having energy $E/N$ when that particle is a part of a larger system of $N$ distinguishable particles. Thus, a single particle of an ideal classical gas may occupy $\Omega(E/N,V,1)$ classical microstates, a system of two such distinguishable particles may occupy $[\Omega(E/N,V,1)]^2$ classical microstates, and a system of $N$ such distinguishable particles may occupy $[\Omega(E/N,V,1)]^N$ classical microstates. Because the particles of an ideal gas composed of distinguishable particles independently occupy their positions in a single-particle phase space, they are said to be statistically independent.

3. Physical Basis of the Average Energy Approximation

We adapt this description of an ideal classical gas of distinguishable particles to the requirements of quantum physics by: (1) replacing the arbitrarily sized unit of action $H$ with Planck's constant $h = 6.63 \cdot 10^{-24} \cdot m^2 kg/s$, (2) giving up the classical fiction that identical particles are always distinguishable, and (3) adopting $\Omega(E/N,V,1)$ as the number of microstates available to a single particle of an N-particle ideal quantum gas when that particle has energy $E/N$ and occupies volume $V$.

Assumption (3) follows from our interpretation of $\Omega(E/N,V,1)$ as the number of single-particle, classical microstates available to a particle of ideal gas composed of distinguishable particles with energy $E/N$ and the insight that *the number of microstates available to a single particle of ideal gas must be independent of whether that particle is part of a larger system of distinguishable or indistinguishable particles, and, if indistinguishable, whether fermions or bosons*. The number of microstates available to a single particle of ideal gas $\Omega(E/N,V,1)$ is independent of the kind of particles that occupy this phase space because $\Omega(E/N,V,1)$ describes the structure of this phase space rather than the kinds of particles that may fill it. The structure is itself common to the particles of any kind of ideal gas.

This identification of the quantum version of $\Omega(E/N,V,1)$ with the number of single-particle microstates available to a particle of any ideal gas when that particle has the average energy $E/N$ is so important to the average energy approximation we adopt the symbol

$$n = V\left(\frac{E}{N}\right)^{3/2}\left(\frac{4\pi em}{3h^2}\right)^{3/2} \tag{4}$$

as more convenient than $\Omega(E/N,V,1)$.

4. The Multiplicity of an Ideal Gas

How many distinct ways can $N$ identical bosons of an ideal Bose-Einstein gas occupy $n$ single-particle microstates? The answer is the multiplicity

$$\Omega(E,V,N) = \frac{(N+n-1)!}{N!(n-1)!} \quad . \tag{5}$$

Alternatively, the number of distinct ways $N$ identical fermions of an ideal Fermi-Dirac gas can occupy $n$ single-particle microstates when no more than one particle may occupy each microstate is the multiplicity

$$\Omega(E,V,N) = \frac{n!}{(n-N)!N!} \quad . \tag{6}$$

Of course, when the particles are assumed distinguishable from each other the multiplicity is given by

$$\Omega(E,V,N) = n^N \quad . \tag{7}$$

In each of these three cases (5)-(7) we have made explicit the dependence of the multiplicity $\Omega$ on the system's extensive variables, $E$, $V$, and $N$, through the dependence of $n$ on these variables as given by (4). It was Einstein's use of the combinatoric formula (5) that prompted Paul Ehrenfest to note that Einstein had abandoned the statistical independence of the particles of an ideal gas implicit in (7) [27].

5. The Entropy

The entropy of an ideal Bose-Einstein gas is given by

$$S = k \ln \Omega$$

$$= k \ln \left[ \frac{(N+n-1)!}{N!(n-1)!} \right]. \tag{8}$$

We use Stirling's approximation to simplify (8) whenever, in addition to $N \gg 1$, we have $n \gg 1$. In this case,

$$S = Nk \left\{ (N+n-1)\ln(N+n-1) - N\ln N - (n-1)\ln(n-1) \right\}$$

$$= Nk \left\{ \left(\frac{n}{N}\right) \ln\left(1 + \frac{N}{n}\right) + \ln\left(1 + \frac{n}{N}\right) \right\}. \tag{9}$$

The entropy (8) or (9) and the functional dependence $n(E,V,N)$ given in (4) completely determine the thermodynamics of an ideal Bose-Einstein gas in the average energy approximation. Its properties, however, remain to be explored.

First, we note that the normalized entropy

$$\frac{S}{Nk} = \left(\frac{n}{N}\right) \ln\left(1 + \frac{N}{n}\right) + \ln\left(1 + \frac{n}{N}\right) \tag{10}$$

depends only on the ratio $n/N$ or $N/n$ where the latter, $N/n$, is the number of particles per single-particle microstate. Because this ratio is a useful parameter, we give it a name, the *occupancy*, and a definition

$$\frac{N}{n} = \left(\frac{N}{V}\right)\left(\frac{N}{E}\right)^{3/2}\left(\frac{3h^2}{4\pi em}\right)^{3/2} \quad (11)$$

consistent with the definition of the number of single-particle microstates $n$ found in equation (4). The occupancy is simply the number of particles in the gas divided by the number of single-particle, average energy microstates available to the particles in the gas. The domain of the occupancy is

$$0 < \frac{N}{n} \leq N \quad (12)$$

although at its upper limit where $n=1$ the Stirling approximation used to generate the entropy (9) or (10) is no longer valid. When it is necessary to describe a macrostate composed of one single-particle microstate, we revert to the combinatoric definition of entropy (8) accordingly to which $S=0$ when $n=1$.

When $N/n \ll 1$ the leading order terms in an expansion of the normalized entropy (10) produce

$$\frac{S}{Nk} = 1 + \ln\left(\frac{n}{N}\right)$$

$$= 1 + \ln\left[\left(\frac{V}{N}\right)\left(\frac{E}{N}\right)^{3/2}\left(\frac{4\pi em}{3h^2}\right)^{3/2}\right]$$

$$= \frac{5}{2} + \ln\left[\left(\frac{V}{N}\right)\left(\frac{E}{N}\right)^{3/2}\left(\frac{4m\pi}{3h^2}\right)^{3/2}\right] \quad (13)$$

where in the first step of this equation sequence we have used the definition of occupancy (11). This result (13) is the Sakur-Tetrode entropy of a monatomic, ideal, classical gas expressed in terms of its extensive parameters. Apparently, the low occupancy limit is the classical limit.

6. Equations of State

Equations of state follow from the derivatives of the entropy (9) with respect to its extensive variables as expressed parametrically by the occupancy (11). Thus, $\left(\partial S/\partial E\right)_{V,N} = 1/T$ and $\left(\partial S/\partial V\right)_{E,N} = P/T$ generate

$$\frac{E}{NkT} = \left(\frac{3}{2}\right)\left(\frac{n}{N}\right)\ln\left(1+\frac{N}{n}\right) \tag{14}$$

and

$$\frac{PV}{NkT} = \left(\frac{n}{N}\right)\ln\left(1+\frac{N}{n}\right) \ . \tag{15}$$

On combining these two we find that

$$E = \frac{3}{2}PV \ . \tag{16}$$

It is easy to see that these equations of state reduce to $E = 3NkT/2$ and $PV = NkT$ in the low occupancy, $N/n \ll 1$, regime. Note that Planck's constant $h$ persists in the equations of state (14) and (15) through its appearance in the definition of the occupancy (11). The equation of state that follows from $\left(\partial S/\partial N\right)_{V,N} = -\mu/T$ where $\mu$ is the chemical potential is not needed in this analysis.

7. Low Temperature Regime

Does the entropy (9) observe the third law of thermodynamics? In other words, does the entropy $S \to 0$ as $T \to 0$? While we cannot analytically solve the entropy (9), the energy equation of state (14), and the occupancy (11) for the function $S(T)$, we can use the occupancy $N/n$ as a parameter that links the normalized entropy $S/Nk$ given by (10) and a normalized temperature, and, in this way, numerically compute and plot $S(T)$. But first we need to discover a way to normalize the temperature.

We find it convenient to use the internal energy $E_o$ achieved when all $N$ bosons occupy the same single-particle microstate. In this case, $n=1$ and $E = E_o$. We also assume in this normalization that the system volume $V$ remains constant. In this case (4) reduces to

$1 = V(E_o/N)^{3/2}(4\pi em/3h^2)^{3/2}$, that is, to

$$E_o = \frac{N}{V^{2/3}}\left(\frac{3h^2}{4\pi em}\right) . \qquad (17)$$

In terms of $E_o$ the definition of the number of single-particle microstates $n$ as given by (4) reduces to

$$n = \left(\frac{E}{E_o}\right)^{3/2} . \qquad (18)$$

Using this relation to eliminate the internal energy $E$ from the energy equation of state (14) produces an expression for the normalized temperature $T$ in terms of the occupancy $N/n$, that is,

$$\frac{N^{1/3}kT}{E_o} = \left(\frac{2}{3}\right)\frac{(N/n)^{1/3}}{\ln(1+N/n)} . \qquad (19)$$

This result (19) allows us to plot, in figure 1, the isochoric normalized entropy $S/Nk$ described by (10) versus the normalized temperature $N^{1/3}kT/E_o$ described by (19). The occupancies that produce figure 1 range from $N/n=1$ (upper right) to $N/n=100$ (lower left-center). The parameters $N$ and $V$ are held constant.

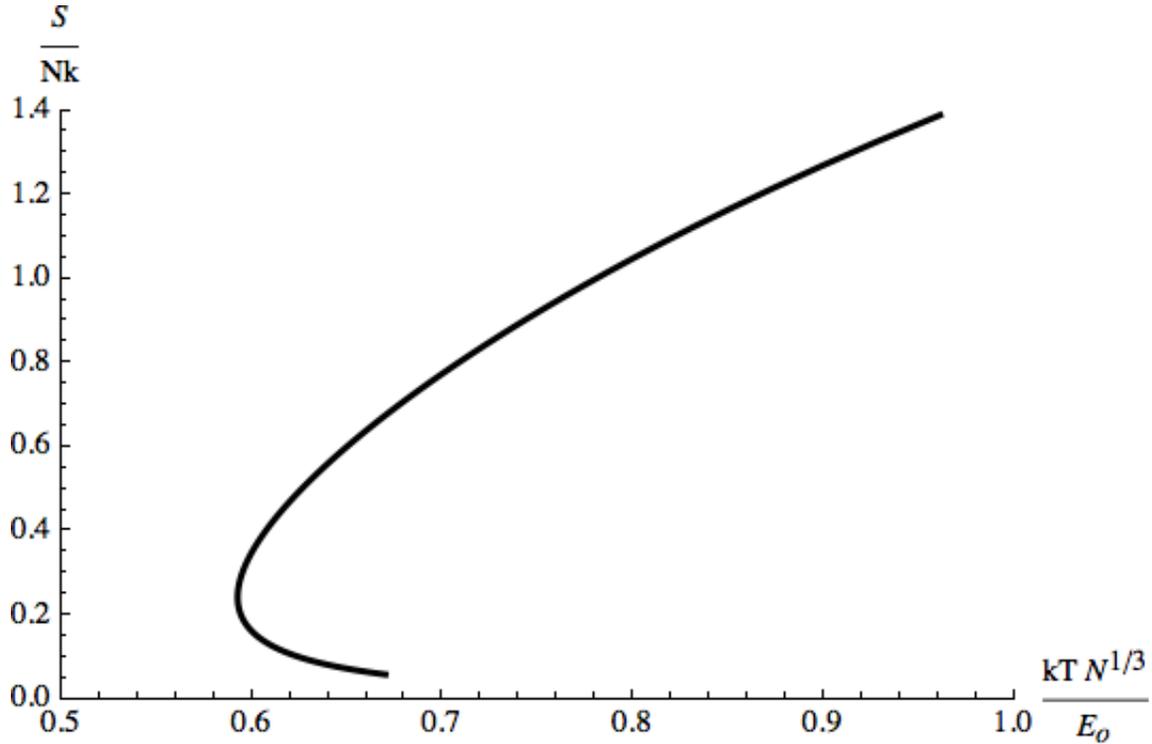

Figure 1. Normalized entropy $S/Nk$ versus normalized temperature $N^{1/3}kT/E_o$ for an ideal Bose-Einstein gas from (10) and (19). The particle number $N$ and volume $V$ are held constant.

According to figure 1, the entropy generally decreases with the temperature, but near $N^{1/3}kT/E_o \approx 0.59$ where $N/n \approx 16$ the derivative $(\partial S/\partial T)_{V,N}$ diverges. When $N/n > 16$ the entropy doubles back and the lower branch of the $S(T)$ curve has a negative slope, that is, a negative $(\partial S/\partial T)_{V,N}$. But when its heat capacity $C_V \left[ = T(\partial S/\partial T)_V \right]$ is negative, a fluid cannot remain in thermal equilibrium with a constant temperature environment. Therefore, this system is dramatically unstable for occupancies $N/n > 16$. In addition, as we shall see, the system is, for a different reason, also thermodynamically unstable for all occupancies larger than a critical occupancy $N/n_c \ [=0.392]$.

8. Thermodynamic Instability

An ideal Bose-Einstein gas in thermal and mechanical equilibrium with an environment that acts as a temperature and pressure reservoir becomes thermodynamically unstable whenever the gas finds itself at a local maximum of its Gibbs free energy $G \ [=E+PV-TS]$. This happens whenever the Gibbs free energy of the gas in an $n >> 1$ macrostate described by the entropy (9) or (10) becomes equal to or greater than the Gibbs free energy of the gas in its $n=1$, $S=0$ macrostate. This condition is

$$G_{n>>1} \geq G_{n=1}$$

$$(E+PV-TS)_{n>>1} \geq (E+PV-TS)_{n=1}$$

$$E_{n>>1} + PV_{n>>1} - TS_{n>>1} \geq E_{n=1} + PV_{n=1} - TS_{n=1}$$

$$E_{n>>1} + PV_{n>>1} - TS_{n>>1} \geq E_o + PV_o - TS_o$$

$$\frac{5}{3} E_{n>>1} - TS_{n>>1} \geq 0 \qquad (20)$$

where we have, on the left hand side, used (14), that is, $E=3PV/2$. The third step in this equation sequence is a matter of the definitions $E_{n=1}=E_o$, $S_{n=1}=S_o$, and $V_{n=1}=V_o$. In the fourth step we use the fact that $S_o=0$ and assume that $E_{n>>1} >> E_o$ and $V_{n>>1} >> V_o$. Condition (18) is equivalent to the chemical potential $\mu_{n>>1}$ of the $n>>1$ state exceeding the chemical potential $\mu_{n=1}$ of the $n=1$ state and arguing that the latter is relatively small.

Using (10) and (14) to eliminate, respectively, the entropy and the energy from the left hand side of (20) produces a condition

$$\left(\frac{3}{2}\right)\left(\frac{n}{N}\right)\ln\left(1+\frac{N}{n}\right) - \ln\left(1+\frac{n}{N}\right) \geq 0 \qquad (21)$$

and a critical occupancy

$$\frac{N}{n_c} = 0.392 \qquad (22)$$

that identifies the onset of thermodynamic instability. For occupancies smaller than this critical occupancy the ideal Bose-Einstein gas is thermodynamically stable and for occupancies greater than this critical occupancy the ideal Bose-Einstein gas is thermodynamically unstable.

This critical occupancy $N/n_c \; [=0.392]$ corresponds to a normalized critical temperature (19) of $N^{1/3}kT_c/E_o = 1.48$ or, equivalently, to a critical temperature

$$T_c = 1.48 \frac{E_o}{N^{1/3}k}$$

$$= 0.130 \left(\frac{N}{V}\right)^{2/3} \left(\frac{h^2}{mk}\right). \qquad (23)$$

Thus the ideal Bose-Einstein gas is stable for temperatures above $T_c$ and unstable for temperatures below $T_c$. This critical point is on the curve shown in figure 1 but beyond the figure's upper right hand corner. Conditions approximating those realized in the first Bose-Einstein condensate of an ideal gas [28], that is, 1000 isotope 87 rubidium atoms contained within a 3-micron radius sphere, yield, according to (23), a critical temperature of 195 nanokelvins.

9. Two Phase Region

When the temperature of an ideal Bose-Einstein gas is cooled below its critical temperature $T_c$ with entropy given by (9) or (10) the gas finds itself at an unstable, local maximum of its Gibbs free energy. Consequently, the gas seeks nearby local minima and in the process splits into two parts. One part of the gas occupies the zero-entropy, $n=1$ macrostate or phase called a Bose-Einstein condensate, while the other part occupies the $N/n_c = 0.392$, critical or marginally stable macrostate or phase.

We can determine the entropy associated with the particles in these two phases in terms of the system temperature $T$ when $0 \leq T \leq T_c$ in the following way. First, let $S_{un}$ stand for the entropy of the $N_{un}$ particles in the $n \gg 1$, uncondensed phase and let $S_o [=0]$ stand for the entropy of the $N_o$ particles in the $n=1$, condensed phase. The relation

$$N_{un} + N_o = N \qquad (24)$$

expresses particle conservation.

The key to determining how the fractions of particles in each phase, $N_{un}/N$ and $N_o/N$, depend on the temperature $T$ in the two-phase region where $0 \leq T \leq T_c$ is to realize that the $N_{un}$ particles in the uncondensed phase remain marginally stable. The relation

$$\frac{N_{un}}{n_{un}} = \frac{N}{n_c}[=0.392] \qquad (25)$$

expresses this criterion.

Using the definition of the occupancy (11) to expand both sides of this stability condition (25) produces

$$\frac{N_{un}}{n_{un}} = \frac{N}{n_c}$$

$$\left(\frac{N_{un}}{V}\right)\left(\frac{N_{un}}{E_{un}}\right)^{3/2}\left(\frac{3h^2}{4\pi em}\right)^{3/2} = \left(\frac{N}{V}\right)\left(\frac{N}{E_c}\right)^{3/2}\left(\frac{3h^2}{4\pi em}\right)^{3/2}$$

$$\frac{N_{un}^{5/2}}{E_{un}^{3/2}} = \frac{N^{5/2}}{E_c^{3/2}} \quad . \tag{26}$$

Similarly, coupling the stability condition (25) with the energy equation of state (14) produces

$$\frac{E_{un}}{N_{un}T} = \frac{E_c}{NT_c} \quad . \tag{27}$$

Eliminating the ratio $E_{un}/E_c$ from (26) and (27) yields

$$\frac{N_{un}}{N} = \left(\frac{T}{T_c}\right)^{3/2} \tag{28}$$

and, given particle conservation (24),

$$\frac{N_o}{N} = 1 - \left(\frac{T}{T_c}\right)^{3/2} \tag{29}$$

when $0 \leq T \leq T_c$. When $T \geq T_c$, there are no particles in the condensed state so that $N_o = 0$. This relation (29) and that describing the critical temperature (23) are, in principle, experimentally verifiable.

We are now in a position to use these fractions, (28) and (29), to compose an expression for the total entropy $S$ of the two-phase system out of the entropies, $S_{un}$ and $S_o$, of its two phases. The normalized entropy in the uncondensed phase is given by

$$\frac{S_{un}}{N_{un}k} = \left(\frac{n_{un}}{N_{un}}\right)\ln\left(1+\frac{N_{un}}{n_{un}}\right) + \ln\left(1+\frac{n_{un}}{N_{un}}\right)$$

$$= \left(\frac{n_c}{N}\right)\ln\left(1+\frac{N}{n_c}\right) + \ln\left(1+\frac{n_c}{N}\right)$$

$$= 2.11 \tag{30}$$

where here we have used the criterion (25). Of course, the entropy of the condensed phase vanishes, that is, $S_o = 0$. Therefore, the total normalized two-phase entropy is

$$\frac{S}{Nk} = 2.11\left(\frac{N_{un}}{N}\right)$$

$$= 2.11\left(\frac{T}{T_c}\right)^{3/2} \tag{31}$$

when $0 \le T \le T_c$.

Figure 2 is similar to figure 1 in that it reproduces the $S(T)$ curve in the region for which $T > T_c$. Below the critical temperature $T_c$, in the two-phase region, the entropy dependence $S(T)$ is determined by (31) and (19). The entropy $S(T)$ shown (solid curve) is continuous at the normalized critical temperature $T_c \left[= 1.48 E_o / N^{1/3} k\right]$. Clearly, this system observes the third law.

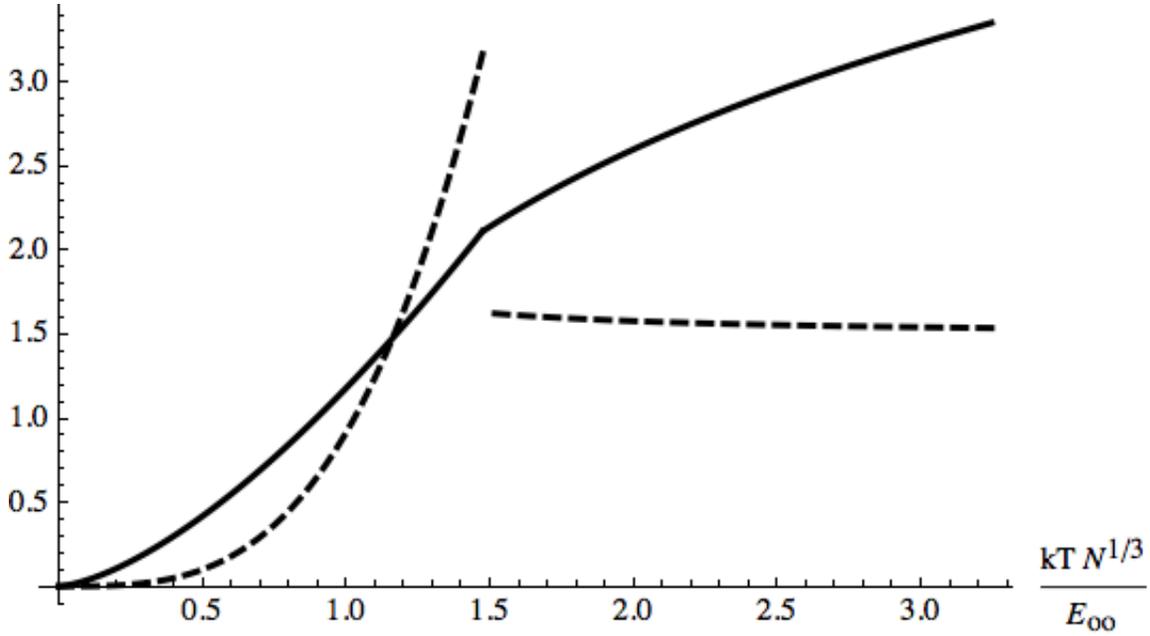

Figure 2. Normalized entropy $S/Nk$ versus normalized temperature $N^{1/3}kT/E_{oo}$ for an ideal Bose-Einstein gas from equations (10) and (19) for $T \geq T_c$ and from (32) and (19) for $T \leq T_c$ (solid). Also the normalized heat capacity $C_V/Nk$ versus normalized temperature from (32) and (19) for $T \geq T_c$ and from equations (33) and (19) for $T \leq T_c$ (dashed).

Also shown in figure 2 is the heat capacity at constant volume $C_V(T)\left[=T(\partial S/\partial T)_{V,N}\right]$. This is derived from $S(T)$ as determined by (10) and (19) when $T \geq T_c$ and from $S(T)$ as determined by (31) and (19) when $0 \leq T \leq T_c$. In this way we find that the normalized heat capacity is

$$\frac{C_V}{Nk} = \frac{(n/N)(1+N/n)\left[\ln(1+N/n)\right]^2}{N/n - (1/3)(1+N/n)\left[\ln(1+N/n)\right]} \tag{32}$$

when $T \geq T_c$ and is

$$\frac{C_V}{Nk} = 2.11 \left(\frac{3}{2}\right)\left(\frac{T}{T_c}\right)^{3/2}$$

$$= 3.16 \left(\frac{T}{T_c}\right)^{3/2} \tag{33}$$

when $0 \leq T \leq T_c$. These heat capacities are plotted as a dashed line in figure 2.

In the low-occupancy, high-temperature regime for which $N/n \ll 1$ the normalized heat capacity (32) reduces, as it should, to $C_V/Nk = 3/2$. The peak heat capacity shown in figure 2, $C_V/Nk = 3.16$, is a little more than twice this classical value. Also the heat capacity vanishes at the absolute zero of temperature just as required by the third law. The continuity of the entropy and the discontinuity of the heat capacity at the critical temperature is a sign of a first-order phase transition.

After investigating the ideal quantum gas that we now call the Bose-Einstein gas Einstein asked Paul Ehrenfest, "The theory [of the ideal Bose-Einstein gas] is pretty, but is there some truth to it?" [29]. The world had to wait some years for experimental verification of Einstein's prediction of a low-temperature condensed phase of an ideal quantum gas. In 1938 Fritz London proposed that the phase transition from liquid to superfluid $^4$He was a transition to a Bose-Einstein condensate. But because $^4$He atoms in a liquid state interact strongly with one another, the condensation is not that of an ideal gas. In 1995 Eric Cornell and Carl Wieman of the University of Colorado cooled a gas of weakly interacting rubidium atoms below the critical temperature and produced a true Bose-Einstein condensate. For this work they, along with Wolfgang Ketterle for shortly thereafter creating a BE condensate with sodium atoms at the Massachusetts Institute of Technology, received the 2001 Nobel Prize in Physics [30,31].

10. Summary and Conclusion

We have used the number of single-particle microstates $n \left[ = V(E/N)^{3/2} (4\pi em/3h^2)^{3/2} \right]$ available to the particles of an ideal gas when those particles each have average energy $E/N$ and the insight that $n$ must be independent of whether the $N$ particles that occupy these states are distinguishable or indistinguishable and if indistinguishable whether fermions or bosons to model the ideal Bose-Einstein gas. The multiplicity of an ideal Bose-Einstein gas is given by $\Omega = (N+n-1)!/[N!(n-1)!]$ and its entropy by $S = k\ln\Omega$. The alternative multiplicity $\Omega = n!/[N!(n-N)!]$ could be used to explore an ideal gas of fermions [32]. The assumption of average energy avoids a number of concepts and techniques required by the standard, probabilistic approach.

The entropy derived reduces to the Sakur-Tetrode entropy in the low occupancy regime and becomes the entropy of an unstable system in the high-occupancy, that is, the $N/n > 0.392$, low-temperature regime. This instability is analyzed with traditional thermodynamic methods – an analysis that describes a two-phase region one of whose phases is the Bose-Einstein condensate. The predicted dependence on the temperature $T$ of the fraction of particles that occupy the condensate in the two-phase region $N_o/N$ reproduces the standard result $N_o/N = 1 - (T/T_c)^{3/2}$, while the critical temperature predicted $T_c \left[ = 0.130 (N/V)^{2/3} (h^2/mk) \right]$ is 50% larger than the critical temperature $T_c \left[ = 0.0839 (N/V)^{2/3} (h^2/mk) \right]$ predicted by the standard, probabilistic theory. Derivatives of the entropy $S$, such as the heat capacity, are less accurate.

Apparently, the average energy approximation avoids the complications of the probabilistic approach at the cost of losing some accuracy. Nevertheless, the average energy approximation produces at least a rough model of the ideal Bose-Einstein gas and condensate that has the virtue of identifying the minimum ingredients necessary to reproduce its basic phenomena: the quantizing of phase space, the indistinguishability of identical particles, and the number of single-particle microstates common to all ideal gases whose particles have average energy. For these reasons the average energy approximation to the ideal Bose-Einstein gas and condensate should become a valuable pedagogical tool.

Acknowledgements

It is a pleasure to acknowledge contributions by Ralph Baierlein, Elizabeth Behrman, Clayton Gearhart, and Rick Shanahan each of whom either read and commented on a draft of this paper or contributed to a discussion of it.